\documentclass{article}
\usepackage{graphicx}
\usepackage{pifont}
\usepackage[a4paper]{geometry}
\usepackage[cmex10]{amsmath}
\usepackage{array}
\usepackage{multirow}
\usepackage{bm}
\usepackage{authblk}
\usepackage{mdwmath}
\usepackage{xcolor,soul,framed}
\usepackage{cite}
\usepackage{eqparbox}
\usepackage{url}
\usepackage{amsmath}
\usepackage{multirow}
\usepackage{amssymb}
\usepackage{textcomp}
\usepackage{bm}
\usepackage{dblfloatfix}
\usepackage{svg}
\usepackage{anyfontsize}

\newgeometry{
    left=2.2cm,    % left margin
    right=2.2cm    % right margin
}

\begin{document}

\title{Optimal Transmission Switching and Busbar Splitting in Hybrid AC/DC Grids}

\author[1,2]{Giacomo Bastianel}
\author[1,2]{Marta Vanin}
\author[1,2]{Dirk Van Hertem}
\author[1,2]{Hakan Ergun}

\affil[1]{{Department of Electrical Engineering, KU Leuven, Leuven, Belgium} \newline}
\affil[2]{{Etch-EnergyVille, Genk, Belgium}}

\date{}

\maketitle

%%%% AC stuff

%% SETS

% nodes
\newcommand{\acnodes}{\mathcal{I}}
% branches
\newcommand{\acbranches}{\mathcal{L}}
%switches
\newcommand{\acswitches}{\mathcal{SW}^{ac}}
%topologies
\newcommand{\actopology}{\mathcal{T}^{ac}}
\newcommand{\actopologyrev}{\mathcal{T}^{ac, rev}}
% switch topologies
\newcommand{\acswitchtopology}{\mathcal{T}^{\text{sw,ac}}}
\newcommand{\acswitchtopologyrev}{\mathcal{T}^{\text{sw}^{\text{ac, rev}}}}
\newcommand{\acZILtopology}{\mathcal{T}^{\text{ZIL,ac}}}

%% VARIABLES

\newcommand{\nodevoltage}{V_i}
\newcommand{\acbranchflow}{S_{lij}}

%%%% DC stuff

%% SETS

% nodes
\newcommand{\dcnodes}{\mathcal{E}}
% branches
\newcommand{\dcbranches}{\mathcal{D}}
%switches
\newcommand{\dcswitches}{\mathcal{SW}^{dc}}
%topologies
\newcommand{\dctopology}{\mathcal{T}^{dc}}
\newcommand{\dctopologyrev}{\mathcal{T}^{dc, rev}}
% switch topologies
\newcommand{\dcswitchtopology}{\mathcal{T}^{\text{sw,dc}}}
\newcommand{\dcswitchtopologyrev}{\mathcal{T}^{\text{sw}^{\text{dc}, rev}}}
\newcommand{\dcZILtopology}{\mathcal{T}^{\text{ZIL,dc}}}

% ac nodes built when performing busbar splitting
\newcommand{\acnodesnew}{\mathcal{I'}}
\newcommand{\acZIL}{\mathcal{S}}
\newcommand{\dcnodesnew}{\mathcal{E'}}
\newcommand{\dcZIL}{\mathcal{Q}}

%% VARIABLES

\newcommand{\dcbranchflow}{P_{def}}

%%%% Loads, generators, converters

%% SETS

\newcommand{\acdcconverters}{\mathcal{C}}

\newcommand{\convertertopology}{\mathcal{T}^{\text{cv}}}

\newcommand{\generators}{ \mathcal{G}}

\newcommand{\loads}{\mathcal{M}}

\newcommand{\dcgenerators}{\mathcal{G}^{\text{dc}}}

\newcommand{\dcloads}{\mathcal{M}^{\text{dc}}}

\newcommand{\genconn}{\mathcal{T}^{\text{gen}}}
\newcommand{\dcgenconn}{\mathcal{T}^{\text{gen,dc}}}

\newcommand{\acloadconn}{\mathcal{T}^{\text{load}}}
\newcommand{\dcloadconn}{\mathcal{T}^{\text{load, dc}}}

%% VARIABLES

\newcommand{\genpower}{ S^g_k }
\newcommand{\acloadpower}{ S^m_k }
\newcommand{\dcloadpower}{ P^m_k }
\newcommand{\converteracpower}{ S^c_l }
\newcommand{\converterdcpower}{ P^{c, dc}_l }

% .... continua tu con le variabili... :)

%\setlength{\parindent}{0mm}
\newcounter{model1} \setcounter{model1}{0}
\newcounter{model2} \setcounter{model2}{0}
\newcounter{model3} \setcounter{model3}{0}
\newcounter{model4} \setcounter{model4}{0}
\newcounter{model5} \setcounter{model5}{0}
\newcounter{model6} \setcounter{model6}{0}
\newcommand{\modelone}[1]{\noindent%
	\refstepcounter{model1}\text{(M1.\arabic{model1})}\\%
}
\newcommand{\modeltwo}[1]{\noindent%
	\refstepcounter{model2}\text{(M2.\arabic{model2})}\\%
}
\newcommand{\modelthree}[1]{\noindent%
	\refstepcounter{model3}\text{(M3.\arabic{model3})}\\%
}

\begin{abstract}
 Driven by global climate goals, an increasing amount of Renewable Energy Sources (RES) is currently being installed worldwide. Especially in the context of offshore wind integration, hybrid AC/DC grids are considered to be the most effective technology to transmit this RES power over long distances. As hybrid AC/DC systems develop, they are expected to become increasingly complex and meshed as the current AC system. Nevertheless, there is still limited literature on how to optimize hybrid AC/DC topologies while minimizing the total power generation cost. For this reason, this paper proposes a methodology to optimize the steady-state switching states of transmission lines and busbar configurations in hybrid AC/DC grids. The proposed optimization model includes optimal transmission switching (OTS) and busbar splitting (BS), which can be applied to both AC and DC parts of hybrid AC/DC grids. To solve the problem, a scalable and exact nonlinear, non-convex model using a big M approach is formulated. In addition, convex relaxations and linear approximations of the model are tested, and their accuracy, feasibility, and optimality are analyzed. The numerical experiments show that a solution to the combined OTS/BS problem can be found in acceptable computation time and that the investigated relaxations and linearisations provide AC feasible results.  
\end{abstract}


\section{Introduction and motivation}
 Due to the European Union's commitment to decarbonize its power system, a massive increase in installed offshore wind capacity is expected in the North Sea by 2050~\cite{Ostend}. %As a result, AC/DC grids will play a critical role in ensuring the secure and reliable operation of the European power system. In this sense,
For this purpose, the construction of several offshore multi-terminal High-Voltage Direct Current (HVDC) grids in the North Sea area has been announced~\cite{ONDP}. Multi-terminal HVDC (MTDC) grids allow for the exchange of power between different DC substations, inherently increasing the system's redundancy compared to having only point-to-point HVDC interconnections. Moreover, the originally separate HVDC grids are expected to become connected, creating a North Sea-wide offshore HVDC grid. In addition, the use of switching units in DC substations has been recently discussed in~\cite{InterOPERA} and will likely play a major role in future expansions of hybrid AC/DC grids. The future offshore HVDC grid will therefore serve a double purpose: i) the transport of offshore wind generation to shore and ii) the exchange of power between North Sea countries. 

 As a result of bulk offshore Renewable Energy Sources (RES) integration and the projected increase in demand due to the electrification of society~\cite{IEA_2023,IEA_2025}, the future hybrid AC/DC grid is expected to experience congestion~\cite{JRC_2024}. Currently, generation redispatch is the most used congestion mitigation measure despite being considerably costly. For example, the congestion and security-related redispatch costs in Germany have been approximately 2.6 bn€ in 2023~\cite{DE_COSTS}, and 2.774 bn€ in 2024~\cite{SMARD}. The need for congestion management actions is only expected to increase with more RES~\cite{JRC_2024}.
 
 Remedial actions are an alternative, inexpensive congestion management measure. For example, 16 European Transmission System Operators (TSOs) use a day-ahead capacity calculation methodology~\cite{nrao} aimed at maximizing the available transfer capacity in their grid. However, this methodology is still in its development phase, and the additional benefits of remedial actions for hybrid AC/DC grids have not been fully quantified yet, due to the lack of appropriate tools. In addition,~\cite{nrao} relies mainly on well-known remedial actions on the borders between existing market zones, without considering the power flows introduced in the system by new HVDC grids. Evaluating the potential of new remedial actions is critical to ensure an efficient integration of such grids in the power system, maximizing their power transfer while minimizing congestion. %As the proposed offshore HVDC grid development plans are recent and there are still no sizeable Multi-Terminal DC (MTDC) grids in Europe, using topological actions to optimize the grid topology in hybrid AC/DC grids is a novel idea. In addition, Optimal Power Flow (OPF) models computing the power transmitted through HVDC links are still recent in the literature~\cite{ergun_optimal_2019,TNEP_Jay} and their more mathematically challenging applications, such as busbar splitting, are yet to be explored.
Therefore, the \textit{main contribution} of this paper is to propose a model representing the mathematical basis for advanced future grid topology optimization tools. By performing Optimal Transmission Switching (OTS) or Busbar Splitting (BS) on both AC and DC substations of a power system for the first time, the optimal grid topology is determined, minimizing the total generation cost, or redispatch cost, while accounting for the physical constraints of the hybrid AC/DC grid as a whole. Furthermore, multiple relaxations and approximations are developed and implemented for a complete analysis of feasibility, optimality, and tractability.

The developed grid topology optimization model can be utilized in the operational context, or as part of cost-benefit analysis methods to compare grid topologies with varying substation configurations. Especially in the context of offshore MTDC grids, the developed models will be crucial to optimizing the system's topology from the planning and operational point of view.

\section{Related work and contributions}\label{sec:literature_review}
\subsection{Optimal Transmission Switching}
The OTS problem aims to switch on/off certain network components in the power grid to achieve reductions in the total generation costs. The concept is well understood and has been common practice in AC grids for years, before being discussed in the scientific literature~\cite{Old_OTS}. Fisher et al. \cite{Fisher2008} formalized it using the DC linearization of the nonlinear non-convex AC-OPF constraints. In addition, Hedman et al. extended the concept with sensitivity analyses~\cite{Hedman_2008} and contingency analyses~\cite{Hedman_2009}, again using the DC linearization. In addition, (almost all) the same authors of~\cite{Hedman_2009} discussed OTS coupled with network topology optimization in~\cite{Hedman_2011}. However, the solutions of the DC-OTS led to infeasibility when applied to the exact, nonlinear non-convex AC formulation \cite{Soroush2014}. Moreover, there is no guarantee that its solution is a lower bound of the AC model, as the DC linearization is an approximation of the AC-OPF problem~\cite{PowerModels2018}. Note that AC infeasibility of DC solutions also occurs for plain OPF problems~\cite{Baker2021}.  

As the AC-OTS problem is a Mixed-Integer Nonlinear Problem (MINLP), its computational effort is considerable for sizeable (test) networks. Thus, several heuristics-based methodologies have been developed to provide AC-feasible results \cite{Soroush2014},\cite{capitanescu_ac_2014,crozier_feasible_2022}. Soroush et al. \cite{Soroush2014} use a DC-OPF-based
line ranking to reduce the operations costs in the exact AC-OPF model. Capitanescu et al. \cite{capitanescu_ac_2014} relax the binary variables and run several nonlinear OPFs, getting acceptable results compared to the MINLP problem while considerably reducing the computational time. A set of standard DC-OPF problems is selected by a greedy algorithm developed by Crozier et al. \cite{crozier_feasible_2022}. However, the lines ``chosen" to be switched off in the DC-OTS problem do not bring the same benefits to the AC-OTS. More recently, Hinneck et al. \cite{Hinneck2023} proposed a methodology including domain-specific knowledge and parallel heuristics that allow testing OTS on larger test cases as Mixed-Integer Linear Programming (MILP) subproblems.

In this paper, a full AC-OTS formulation for AC/DC grids is introduced. Both AC (branches) and DC (DC branches and AC/DC converters) switching actions can occur, representing the first time that OTS has been implemented and applied to DC grids for a deterministic problem. 

The results presented in Section~\ref{sec:results} use OTS for small to medium-sized hybrid AC / DC grids. We acknowledge the additional challenges that solving the AC-OTS problem for larger test cases would entail. However, the focus and main contribution of the paper is the BS model for hybrid AC/DC grids, implemented in different power flow formulations.

\subsection{Busbar splitting}
The busbar splitting (BS) problem aims to split selected busbars into two (or more) parts to increase the transmission capacity and relieve congestion in a power grid. Intuitively, splitting a busbar increases the electrical distance between its parts, while they remain physically close. BS is mathematically modeled by Heidarifar et al. \cite{Heidarifar2014} to alleviate congestion while satisfying the N-1 security criterion for a given network. An improved node-breaker model is introduced in \cite{Heidarifar2016}, by (largely) the same authors. There, each substation is modeled in its double-busbar configuration. The two ``splittable" parts of the busbar are linked by a Zero Impedance Line (ZIL), and each grid element can be connected to either of them using one binary variable. %ZILs are used as well in our busbar splitting model to link the parts of the busbar being split.
 In our work, the busbar couplers are modeled as ZILs, too.
% As stated in \cite{Morsy2022}, one limitation of \cite{Heidarifar2021} is that every substation can be split into a maximum of two parts. Additionally
The approach in~\cite{Heidarifar2016} was further extended in~\cite{Heidarifar2021} with a second-order cone (SOC) formulation~\cite{SOC}, and N-1 contingency constraints. The substation model in \cite{Heidarifar2021} can represent both a double-bus double-breaker and a breaker-and-a-half arrangement. Nevertheless, the method in~\cite{Heidarifar2021} only allows a limited set of switching actions and may not be sufficiently expressive to adequately describe all feasible busbar configurations, as busbar switching actions are only described by a single binary variable. The optimization models proposed in~\cite{Heidarifar2014,Heidarifar2016,Heidarifar2021} incorporate N-1 constraints, but feature simpler BS models. In contrast, the main contribution of our paper is to propose a grid topology optimization model with busbar splitting in hybrid AC/DC grids for several formulations, testing whether the results are feasible and optimal for AC-OPF formulations. Future work will include N-1 constraints in the proposed optimization model. Similarly, Hinneck et al.~\cite{Hinneck2021} and Morsy et al.~\cite{Morsy2022} use an augmented network representation to represent the possible switching actions in the mathematical model of a substation. In the augmented network representation, each grid element (generator, load, branch, etc.) is linked to either part of the split busbar through provisional auxiliary branches represented by binary variables. The solver thus optimizes the busbar topology by selecting one of the two binary variables and the state of the busbar coupler, splitting the selected busbar. To overcome the limitations identified in the afore-mentioned work, this paper extends the method in~\cite{Hinneck2021,Morsy2022} by adding AC and DC switches instead of auxiliary branches. This offers the possibility to model protection devices in more detail, e.g., with their specified thermal ratings for both AC and DC substations. In addition, Morsy et al.~\cite{Morsy2022} only consider the DC linearisation of the BS problem, and Heidarifar et al.~\cite{Heidarifar2021} use a SOC relaxation. In this paper, we explore multiple formulations of the AC power flow constraint, including the exact one, which results in a MINLP. Specifically, we test relaxations based on SOC~\cite{SOC}, and convex-quadratic~\cite{QC_math} (QC) programming. The QC has proved valuable for OTS in~\cite{QC}, where it provides accurate lower bounds to the original AC-OTS. Additionally, the ``LPAC'' piecewise-linear approximation from~\cite{LPAC} is modeled\textcolor{blue}{\footnote{The LPAC formulation used in this paper is based on its implementation as a convex-quadratic model in PowerModels.jl~\cite{PowerModels2018} for computational efficiency.}}. Contrary to the DC linearization, it preserves reactive power and voltage magnitudes, implying a better physical representation of the grid. Additionally, a big M reformulation~\cite{Linear_optimization} of the MINLP problem and reformulations are used to linearize all bilinear constraints in the formulation, as in~\cite{Heidarifar2016,Heidarifar2021,BS_wind}. Finally, the results with the different reformulations are compared and checked for AC feasibility. 

Recent work employs BS for voltage stability under contingencies \cite{BSvoltage2023}, distribution network reconfiguration for reactive power regulation with electric vehicles \cite{EV2023}, and to minimize the additional operating expenses required to mitigate the adverse effects of contingencies~\cite{Demarco2023}. Neither these references nor the rest of the literature address the use of remedial actions for DC grids (let alone combined hybrid AC/DC) due to the novelty of the topic and the added complexity of modeling the DC components, such as the AC/DC converter and the HVDC cable poles. Finding reliable AC/DC test cases is still an open challenge, too. Furthermore, the mentioned references resort to simplified models of switching elements, which are represented through binary variables, ignoring their nature as self-standing grid elements with their own thermal limits. 

\subsection{Contributions to the state of the art}
Table~\ref{table:ref} provides an overview of the literature on grid remedial actions and highlights this paper's contributions to the topic.
The paper presents a model that utilizes the full flexibility of remedial actions in hybrid AC/DC grids. This work improves on the state of the art by making the following original contributions:

\begin{enumerate}
    \item A full non-convex steady-state MINLP formulation is developed and implemented for both optimal transmission switching and busbar splitting. 
    \item Optimal transmission switching and busbar splitting are formulated for the DC part of hybrid AC/DC grids for the first time.
    \item Multiple optimal power flow formulations are implemented in a grid topology optimization model, ranging from an exact, nonlinear, non-convex, and mixed-integer formulation to a mixed-integer piecewise linear.
    \item Optimal transmission switching and busbar splitting are combined in a single model for hybrid AC/DC grids. The model allows addressing hybrid AC/DC grids as a whole, evaluating the optimal actions across all AC- and DC-side possibilities in the selected busbars.
    %\item The developed models are made open-source and publicly available in the PowerModelsTopologicalActions Julia package.
\end{enumerate}

\begin{table*}[h!]
\caption{Overview of remedial actions literature. \textbf{\ding{51}} refers to this paper.}
\centering
{\fontsize{7.5pt}{11pt}\selectfont
\begin{tabular}{l|ccc|ccc|ccc}
\hline
  & \multicolumn{3}{c|}{Busbar} & \multicolumn{3}{c|}{Optimal} & \multicolumn{3}{c}{Busbar Splitting with}     \\
    & \multicolumn{3}{c|}{Splitting} & \multicolumn{3}{c|}{Transmission Switching} & \multicolumn{3}{c}{Optimal Transmission Switching}     \\
\cline{2-10}
\multirow{-3}{*}{Formulation} & AC & DC & AC/DC & AC& DC & AC/DC & AC & DC & AC/DC \\
\hline
DC linearization - with binaries      &  \cite{crozier_feasible_2022,Heidarifar2014,Morsy2022,BS_wind}         &       -    &     -         & \cite{Fisher2008,Hedman_2008,Hedman_2009,Hedman_2011,Soroush2014}      &   -    &    -     &  ~\cite{Heidarifar2016}       &     -        &        -     \\
%DC linearization - binaries relaxed   &           &           &              &       &       &         &             &             &             \\
LPAC linearization - with binaries    &   \textbf{\ding{51}}        &   \textbf{\ding{51}}        &     \textbf{\ding{51}}         &  \cite{Brown_LPAC_OTS}     &   -    &    -     & \textbf{\ding{51}} & \textbf{\ding{51}} & \textbf{\ding{51}} \\
%LPAC linearization - binaries relaxed &           &           &              &       &       &         &             &             &             \\
SOC relaxation - with binaries        &   ~\cite{Heidarifar2021},~\textbf{\ding{51}}        &   \textbf{\ding{51}}        &     \textbf{\ding{51}}         &   -    &    -   &   -      & ~\cite{Heidarifar2021},~\textbf{\ding{51}} & \textbf{\ding{51}} & \textbf{\ding{51}} \\
%SOC relaxation - binaries relaxed     &           &           &              &       &       &         &             &             &             \\
QC relaxation - with binaries         &   \textbf{\ding{51}}        &    \textbf{\ding{51}}       &     \textbf{\ding{51}}         &  \cite{QC}     &       &         & \textbf{\ding{51}} & \textbf{\ding{51}} & \textbf{\ding{51}} \\
%QC relaxation - binaries relaxed      &           &           &              &       &       &         &             &             &             \\
Exact, non-convex - with binaries     &    \textbf{\ding{51}}       &  \textbf{\ding{51}}        &    \textbf{\ding{51}}          &   \cite{capitanescu_ac_2014}, \cite{QC}, \textbf{\ding{51}}    &  \textbf{\ding{51}}     &    \textbf{\ding{51}}     & \textbf{\ding{51}} & \textbf{\ding{51}} & \textbf{\ding{51}} \\
Exact, non-convex - binaries relaxed  &       -    &      -     &       -      &   \cite{capitanescu_ac_2014}    &    -   &    -     &       -      &        -     &     -        \\
\hline
\end{tabular}
}
\label{table:ref}
\end{table*}

\section{Methdology}
This work extends the exact, non-convex OPF but continuous formulation for AC/DC grids from the PowerModelsACDC Julia package~\cite{ergun_optimal_2019} by introducing additional variables and constraints to incorporate optimal remedial actions. Thus, this Section describes the full OTS and BS MINLP optimization model.

% \subsection{Nomenclature}

%%%%%%%%%%%%
\subsection{Sets}
%%%%%%%%%%%%
The list of network components, topologies, connectivities, and nomenclature to unambiguously distinguish each network component is presented below. By using `reverse' topologies for AC and DC branches, the same branches described before can represent flows in the opposite direction. Note that both radial and meshed topologies (both on the AC and DC sides) are supported by the proposed model.
\begin{table}[!ht]
	% increase table row spacing, adjust to taste
	\renewcommand{\arraystretch}{1.1}
	\centering
	\label{tb:Model1}
        {\fontsize{8pt}{11pt}\selectfont
	\begin{tabular}{m{20em} l}
		\hline
		Network components &\\
		\hline
            $\acnodes$          & Set of AC nodes             \\           
            $ \acbranches $     & Set of AC branches  \\          
            $ \acswitches $     & Set of AC switches  \\          
            $ \dcnodes $        & Set of DC nodes        \\          
            $ \dcbranches $     & Set of DC branches  \\          
            $ \dcswitches $     & Set of DC switches  \\          
            $ \acdcconverters $ & Set of AC/DC converters  \\
            $ \generators $     & Set of AC generators   \\
            %$ \dcgenerators $   & Set of DC generators \\
            $ \loads $          & Set of AC loads   \\
            %$ \dcloads $        & Set of DC loads  \\
            $ \acnodesnew $     & Set of AC nodes for BS \\
            $ \acZIL $          & Set of AC switches for BS \\
            $ \dcnodesnew $     & Set of DC nodes for BS  \\
            $ \dcZIL $          & Set of DC switches for BS \\
 		\hline
		Topologies and connectivities &\\
		\hline
            $\actopology$         & AC topologies             \\
            $\actopologyrev$      & Reverse AC topologies     \\ 
            $ \dctopology $       & DC topologies             \\
            $ \dctopologyrev$     & Reverse DC topologies     \\
            $ \acswitchtopology $ & AC switch topologies      \\
            $ \dcswitchtopology $ & DC switch topologies      \\
            $ \acZILtopology $    & AC busbar couplers   \\    
            $ \dcZILtopology $    & DC busbar couplers   \\   
            $ \convertertopology$ & AC/DC converter topologies\\
            $ \acloadconn $       & AC load connectivity      \\
            %$ \dcloadconn $       & DC load connectivity      \\
            $ \genconn $          & AC generator connectivity \\
            %$ \dcgenconn $        & DC generator connectivity \\
            \hline
		Nomenclature of each network component &\\
            \hline
            $lij~\in~\actopology~\subseteq~\acbranches~\times~\acnodes~\times~\acnodes$ & AC branches \\
            $\upsilon ii'~\in~\acZILtopology \subseteq~\acswitches~\times~\acnodes~\times~\acnodesnew$ & AC bus couplers \\
            $\upsilon mi~\in~\acswitchtopology\subseteq~\acswitches~\times~\acnodes~\times~\acnodesnew$ & AC switches \\
            $def~\in~\dctopology~\subseteq~\dcbranches~\times~\dcnodes~\times~\dcnodes$ & DC branches \\ 
            $\xi ee'~\in~\dcZILtopology~\subseteq~\dcswitches~\times~\dcnodes~\times~\dcnodesnew$ & DC bus couplers \\
            $\xi re~\in~\dcswitchtopology~\subseteq~\dcswitches~\times~\dcnodes~\times~\dcnodesnew$ & DC switches \\
            $lji~\in~\actopologyrev~\subseteq~\acbranches~\times~\acnodes~\times~\acnodes$ & Reverse AC branches \\
            $dfe~\in~\dctopologyrev~\subseteq~\dcbranches~\times~\dcnodes~\times~\dcnodes$ & Reverve DC branches \\

            $mi~\in~\acloadconn$ & Loads \\
            $gi~\in~\genconn$ & Generators \\
            $cie~\in~\convertertopology$ & AC/DC Converters \\
            
            \hline
	\end{tabular}
    }
\end{table}

\subsection{Parameters}\label{sec:parameters}
All the generators $k$ in the network are assigned cost parameters $c_{2k}$ (\$/$W^{2}$), $c_{1k}$ (\$/W),  $c_{0k}$ (\$). Note that the quadratic cost term of each generator is assigned to zero. Each converter $c$ has loss parameters  {$a_{cv} \geq 0$ (W), $b_{cv} \geq 0$ (W/A), $c_{cv} \geq 0$ ($\Omega$). The values used in the model are taken from Ergun et al.~\cite{ergun_optimal_2019}, where: ``\textit{$a_{cv}$ represents the no-load losses of transformers and averaged auxiliary equipment losses, $b_{cv}$ represents the switching losses of valves and freewheeling diodes, while $c_{cv}$ is related to the conduction losses of the valves}''.  
A typical value for $a_{cv}$ is $11.033 \cdot 10^{-3}$ pu, for $b_{cv}$ is $3.464 \cdot 10^{-3}$ pu, for $c_{cv}$ is $4.4 \cdot 10^{-3}$ pu for the rectifying and $6.6 \cdot 10^{-3}$ pu for the inverting converter have been used in the literature~\cite{30PMACDC,31PMACDC}.

A generic HVDC converter station model is used, whose parameters and (continuous) variables are shown in Fig.~\ref{fig:Converter_station}. Its transformer and phase reactor are characterized by the admittances $y^{tf}_c = (z^{tf}_c)^{-1} = g_c^{tf} + j b_c^{tf}, y^{pr}_c = (z^{pr}_c)^{-1} = g_c^{pr} + j b_c^{pr}$. The filter has a shunt capacitor with susceptance $b^f_c$. Finally, $t_c$ indicates the transformer voltage magnitude transformation factor.

\begin{figure}[h!]
		\centering
		\includegraphics[scale=0.75]{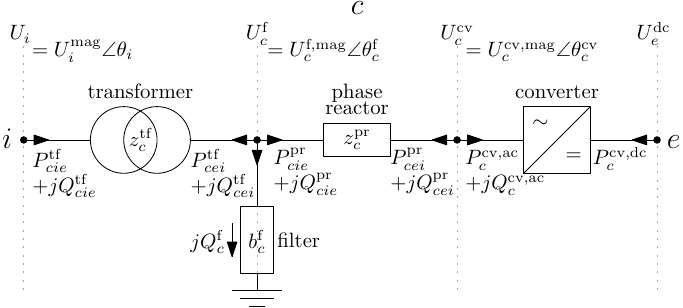}
		\caption{Converter station model, figure  from~\cite{ergun_optimal_2019}.} 
		\label{fig:Converter_station}
\end{figure}
Each DC branch $d$ is modeled as a series resistance $r_d$ (or conductance $g_d = r_d^{-1}$, left-hand side of Fig.~\ref{fig:branch_model}), whereas AC branches $l$ are characterized by a $\pi$-model (right-hand side of Fig.~\ref{fig:branch_model}) with a series admittance, $y_l = g_l + j b_l$, and two shunt ones, i.e. $y_i, y_j$. All impedance parameters are given as input\footnotemark[2]. Note that load data are input parameters for each test case. 

\begin{figure}[t]
		\centering
\includegraphics[width=.7\textwidth]{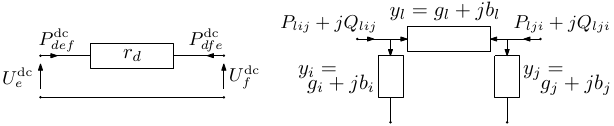}
		\caption{DC (left) and AC (right) branch model.} 
		\label{fig:branch_model}
\end{figure}

Decision variables are defined for generator setpoints, active and reactive power through AC branches, AC voltage angles and magnitudes, AC/DC converter setpoints, power through DC branches, and DC voltage magnitudes. Model 1 includes the variables in the AC-OPF formulation for hybrid AC/DC grids. All variables $X$ are assigned upper and lower bounds, indicated with over- ($\overline{X}$) and underlines ($\underline{X}$), respectively. %These correspond to limits such as rated capacities and operational voltage limits.

\footnotetext[2]{See \cite{ergun_optimal_2019} for a thorough description of all the parameters being mentioned in this section.}

\subsection{Remedial actions models} \label{sec:OPF}
\subsubsection{AC-OPF model for hybrid AC/DC grids}
The AC-OPF model is based on the formulation for hybrid AC/DC grids from \cite{ergun_optimal_2019}. To avoid redundancy in the text, the compact formulation of the OPF model is introduced by (M1.1)-(M1.17).
While our implementation is easily extendable to line-commutated converters (LCC), this paper focuses on voltage-sourced converter (VSC) based HVDC, as this configuration is the standard for new offshore wind connections and is the most suitable technology for multi-terminal DC grids. 
The variable space depends on the chosen underlying power flow formulation. For the AC part of the grid, we use the non-convex AC polar formulation, which is a bus injection model whose variables include apparent power flows/injections $S_{lij},S_{lji}$ at all AC branches/injection buses, and nodal voltage phasors in polar coordinates: $U_i~=~U_{i}^{m}~\angle~\theta_i $. The DC part of the grid is modeled with its exact quadratic formulation. The DC-related variables consist of real power flows/injections at all DC branches/injection buses $P^{dc}_{def}$ and DC voltage magnitude at all nodes $U^{\text{dc}}_e$.

The objective function in (M1.1) consists of minimizing the total costs of the $ \generators $ set of generators,
where $P_{k}^{g}$ is the active power injection. (M1.2) sets the voltage angle of the reference bus to zero while (M1.3) and (M1.4) limit respectively the voltage magnitude and angles of the other buses between a minimum $\underline{U}^{m}_{i},\underline{\theta}_{i}$ and maximum $\overline{U}^{m}_{i},\overline{\theta}_{i}$ value. Similarly, the generator setpoints are bounded by $\underline{S}^{g}_{k},\overline{S}^{g}_{k}$ in (M1.6). Moreover, the AC power balance is shown by (M1.5), where $S^{cv,ac}_c$
refers to injections from HVDC converters, and $S^m_l$ to the
nodal demand. The term $y_{i} \cdot |U_i|^2$ refers to the power absorbed by shunt elements connected to the node $i$. They are considered negligible in the remainder of the paper. Ohm’s law is satisfied using (M1.7) and (M1.8), which refer to the $\pi$-section model for AC branches in Fig. \ref{fig:branch_model}. $Y^*_{lij}$ represents the admittance value in the admittance matrix of the selected network element, while the complex tap ratio $T_{lij}$ is assumed to be unitary, and $\bf{j}$ refers to the imaginary operator. The power flow through the AC branches is constrained by (M1.9), while the difference in the voltage angles across AC branches is bounded by $[{\theta}^{\Delta_{min}}_{lij},{\theta}^{\Delta_{max}}_{lij}] = [-\frac{\pi}{6},\frac{\pi}{6}]$ in (M1.10). 
The apparent power through the converter AC side is limited between the bounds $\underline{S}^{cv,ac}_c$ and $\overline{S}^{cv,ac}_c$ by (M1.11).

Regarding the DC side of the hybrid AC/DC grid, the power balance for the DC nodes $e$ in $ \dcnodes $ is defined by (M1.15). Similarly to the power balance for the AC side in (M1.5), $P^{g,dc}_k$ refers to DC generators, $P^{c, dc}$ to the DC side of the AC/DC converters, $P^{d,dc}_e$ to the DC loads, $y^s_{e} \cdot |U_e|^2$ to the shunt elements and $P^{dc}_{def}$ to the DC branches. The power flow through the DC branches is regulated by (M1.16), where $p^{dc}_{def}$ is the number of poles. $p^{dc}_{def}$ equals to 2 for bipolar and symmetrical monopolar configurations, and $p^{dc}_{def}$ = 1 for monopolar configurations. 

The DC branches are modeled with a single-line representation, which enables the correct representation of monopolar and balanced bipolar (by halving the power flows) DC links. The extension to unbalanced bipolar configurations is left for future work~\cite{Jat2024}. The power flow through the DC branches is bounded within $[\underline{P}^{dc}_{def},\overline{P}^{dc}_{def}]$ by (M1.17). Finally, the AC/DC converters' AC and DC sides are linked by (M1.12), expressing the losses of the converter dependent on the AC side converter current $I^{cv}_c$, and the parameters $a_{cv}$, $b_{cv}$, and $c_{cv}$. The AC/DC converter apparent power $S^{cv,ac}_c$ on the DC side is related to the converter current $I^{cv}_c$ using the nodal voltage magnitude of the AC grid $U^{m}_{i}$ in (M1.13). In the full formulation of the OPF model from \cite{ergun_optimal_2019}, the converter transformer and filters have been modeled as shown in Fig. \ref{fig:Converter_station}. As these elements can be modeled as AC branches without loss of generality, they are not further described in this paper.

\begin{table}[h!]
	% increase table row spacing, adjust to taste
	\renewcommand{\arraystretch}{1.0}
	\centering
	%		\caption{Nonlinear Nonconvex  grid TNEP problem}
	\label{tb:Model1}
        {\fontsize{8pt}{10pt}\selectfont
	\begin{tabular}{m{38em} l}
		\hline
		Model 1: AC-OPF for hybrid AC/DC grids &\\
		\hline
		\textbf{Minimize:}\\
		 $\sum_{k \in G} c_{2k}\cdot {P_{k}^{g}}^{2} + c_{1k}\cdot P_{k}^{g} + c_{0k}$ & \modelone \\
            \\
         \textbf{Subject to}: & \\
		\textbf{AC bus:}\\		
          $\theta_{r} = 0 $ \label{eq:voltage} & \modelone \\
          $\underline{U}^{m}_{i} \leq U^{m}_{i} \leq \overline{U}^{m}_{i}, \quad \forall i \in \acnodes $ \label{eq:voltage_magnitudes} & \modelone\\  
          $ \underline{\theta}_{i} \leq \theta_{i} \leq \overline{\theta}_{i}, \quad \forall i \in \acnodes$ \label{eq:voltage_angles} & \modelone \\
          $\sum_{\substack{k \in \generators_i}} S^g_k +  \sum_{\substack{lij\in \actopology}} S^{ac}_{lij} + \sum_{\substack{c \in \acdcconverters_{i}}} S^{cv,ac}_c  - y_{i} |U_i|^2 = 
          \sum_{\substack{l \in \loads_{i}}} S^m_l, \quad \forall i \in \acnodes \label{eq:ac_power_balance}$ & \modelone\\
          \\
          \textbf{Generator}&\\
          $\underline{S}^{g}_{k} \leq S^g_k \leq  \overline{S}^{g}_{k}, \quad \forall k \in \generators \label{eq:gen_limit}$ & \modelone\\
          \\
          \textbf{AC branch}&\\
        $S_{lij} = (Y^*_{lij} - \textbf{j} \frac{{b^c}_{lij}}{2}) \cdot  \frac{|U^{m}_i|^2}{|{T}_{lij}|^2} - Y^*_{lij} \cdot \frac{U^{m}_i \cdot U^{m*}_{j}}{{T}_{lij}}, \quad \forall lij\in \actopology $ \label{eq:Sij} & \modelone \\
        $S_{lji} = (Y^*_{lij} -  \textbf{j} \frac{{b^c}_{lji}}{2}) \cdot  \frac{|U^{m}_j|^2}{|{T}_{lij}|^2} -  Y^*_{lij} \cdot  \frac{U^{m*}_i \cdot U^{m}_j}{{T}^*_{lij}}, \quad \forall lji \in \actopology \label{eq:Sji}$& \modelone\\
        $|S_{lij}| \leq  \overline{S}_{lij}, \quad \forall lij \in \actopology \cup \actopologyrev \label{eq:Sijlimit}$& \modelone\\
        $\underline{{\theta}^{\Delta}}_{lij} \leq \angle (U^{m}_i \cdot U^{m*}_j) \leq  \overline{{\theta}^{\Delta}}_{lij}, \quad \forall lij \in \actopology \label{eq:voltage_difference}$& \modelone\\
        \\
        \textbf{AC/DC converter}&\\
        $\underline{S}^{cv,ac}_c \leq S^{cv,ac}_c \leq  \overline{S}^{cv,ac}_c, \;\; \forall c \in \acdcconverters$ & \label{eq:converter_limits} \modelone\\ 
        $ P^{cv,ac}_c + P^{cv, dc}_c = a_{cv} + b_{cv} \cdot |I^{cv,dc}_c| + c_{cv} \cdot |I^{cv,dc}_c|^2, \quad \;\; \forall c \in \acdcconverters$ & \label{eq:convloss} \modelone\\
        $|U^{m}_i|^2  \cdot |I^{cv,dc}_c|^2 = (S^{cv,ac}_c)^2, \quad \;\; \forall c \in \acdcconverters, \quad \;\; \forall i \in \acnodes$ \label{eq:final} & \modelone\\
        \\
        \textbf{DC bus}&\\
        $\underline{U}^{dc}_{e} \leq U^{dc}_{e} \leq \overline{U}^{dc}_{e}, \quad \forall e \in \dcnodes $ \label{eq:dc_voltage_magnitudes} & \modelone\\  
        $\sum_{\substack{k \in \dcgenerators_e}} P^{g,dc}_k + \sum_{\substack{c \in \acdcconverters_e}} P^{cv, dc}_c - \sum_{\substack{l \in \dcloads}} {P^{m,dc}_l} - y_{e} \cdot |U^{dc}_e|^2 =  \sum_{\substack{def} \in \dctopology} P^{dc}_{def}, \; \; \quad \forall e \in \dcnodes \label{eq:dc_power_balance}$ & \modelone \\
        \\
        \textbf{DC branch}&\\
        $P^{dc}_{def} =  p^{dc}_{def}  \cdot Y_{def} \cdot( (U^{dc}_e)^2 - U^{dc}_e \cdot U^{dc}_f), \; \; \quad \forall def \in \dctopology \cup \dctopologyrev$ & \label{eq:Peflow}  \modelone \\
        $\underline{P}^{dc}_{def} \leq P^{dc}_{def} \leq \overline{P}^{dc}_{def}, \;\; \forall def \in \dctopology \cup \dctopologyrev$ & \label{eq:Peflimit}  \modelone\\
    
        \hline
	\end{tabular}
    }
\end{table}

\subsubsection{AC and DC Optimal Transmission Switching model}
The AC-OTS model for hybrid AC/DC grids extends the OPF formulation by including (M2.1)-(M2.13) to determine the optimal switching states (open/closed) of all the branches and AC/DC converters in the grid. Each branch and AC/DC converter is assigned a binary variable that equals $0$ if it is disconnected, and $1$ otherwise. The switching decisions are taken through the binary variables $z^\text{ac}_{l}$ for AC branches, $z^{\text{dc}}_{d}$ for DC branches, and $z^{\text{cv}}_{c}$ for converters. For simplicity, we assume that all branches and converters can be switched. Selecting a subset is of course possible and suggested to avoid having excess binary variables.

Therefore, the active and reactive power parts from the apparent powers $S_{lij}$, $S_{lji}$ in (M1.7)-(M1.8) are written in (M2.1)-(M2.6) in their OTS form. %As anticipated before in the section, the binary variable $z^{ac}_{l}$ represents the switching decision for the AC branch. To check all the line parameters in the equations, the reader is referred to the previous Fig. \ref{fig:branch_model}. 
On the DC side, the DC branches are switched on/off by the binary variable $z^{{dc}}_{d}$ in (M2.7), (M2.11) while the AC/DC converters are linked to $z^{{cv}}_{c}$ in (M2.8). 
The combination of (M2.8) and (M2.9) ensures that the converter losses are zero when the converter is de-energized ($z^{cv}_c = 0$). Similarly, all the parameters in the converter power balance, i.e. the converter's DC current $\overline{I}_c^{cv,m}$ (M2.10), AC active power $P_c^{cv,ac}$ (M2.12) and DC active power $P_c^{cv,dc}$ (M2.13) are zero with $z^{cv}_c = 0$.
Using this formulation, it is possible to set up different OTS problems, e.g., fixing all $z^{ac}$ to 1, and only performing DC OTS. Fixing all binaries to 1 results in a conventional AC-OPF problem.

The main contributions of this paper rest in the models for busbar splitting. Hence, relaxations and approximations for the OTS problem are not discussed, although the rationale for applying them to the studied test cases would be the same. 

\begin{table}[h!]
	% increase table row spacing, adjust to taste
	\renewcommand{\arraystretch}{1.0}
	\centering
	%		\caption{Nonlinear Nonconvex  grid TNEP problem}
	\label{tb:Model211}
        {\fontsize{8pt}{11pt}\selectfont
	\begin{tabular}{m{40em} l}
		\hline
		Model 2.1.1: Optimal Transmission Switching &\\
		\hline
        \textbf{AC Optimal Transmission Switching}&\\
        $ P_{lij} = z_{l}^{ac} \cdot ((g_{i}+g_{l}) \cdot (U^{m}_{i})^2 - g_{l} \cdot U^{m}_{i} \cdot U^{m}_{j}  \cdot \cos(\theta_{i}-\theta_{j}) - b_{i} \cdot U^{m}_{i} \cdot U^{m}_{j}  \cdot \sin(\theta_{i}-\theta_{j})),  \quad \;\; \forall lij \in \actopology$& \label{P_ij} \modeltwo \\
        $ Q_{lij}=z_{l}^{ac} \cdot (-(b_{i}+b_{l}) \cdot (U^{m}_{i})^2+b_{l} \cdot U^{m}_{i} \cdot U^{m}_{j}  \cdot \cos(\theta_{i}-\theta_{j}) - g_{i} \cdot U^{m}_{i} \cdot U^{m}_{j}  \cdot \sin(\theta_{i}-\theta_{j})),  \quad \;\;  \forall lij \in \actopology$ & \label{Q_ij} \modeltwo \\
        $P_{lji}=z_{l}^{ac} \cdot ((g_{i}+g_{l}) \cdot (U^{m}_{j})^2-g_{l} \cdot U^{m}_{j} \cdot U^{m}_{i} \cdot \cos(\theta_{j}-\theta_{i}) - b_{lji} \cdot U^{m}_{j} \cdot U^{m}_{i} \cdot \sin(\theta_{j}-\theta_{i})),  \quad \;\;  \forall lji \in \actopology$ & \label{P_lji} \modeltwo \\
        $Q_{lji}=z_{l}^{ac} \cdot ( -(b_{i}+b_{l})\cdot (U^{m}_{j})^2 +b_{l} \cdot U^{m}_{j} \cdot U^{m}_{i} \cdot \cos(\theta_{j}-\theta_{i}) - g_{l} \cdot U^{m}_{j} \cdot U^{m}_{i} \cdot \sin (\theta_{j}-\theta_{i})),  \quad \forall lij \in \actopology$ & \label{Q_ji} \modeltwo\\
        $ (P_{lij}^2 + Q_{lij}^2) \leq z_{l}^{ac} \cdot (\overline{S}_{lij})^2,  \quad \forall lij \in \actopology \cup \actopologyrev$, \label{PQSij} & \modeltwo \\
        $ - z_{l}^{ac} \cdot \underline{\Delta\theta_i} \leqslant \left(\theta_i-\theta_j\right) \leqslant z_{l}^{ac} \cdot \overline{\Delta\theta_i},  \quad \forall lij \in \actopology \cup \actopologyrev$ \label{voltage_angles} & \modeltwo \\ 
        \\
        \textbf{DC Optimal Transmission Switching}&\\
        $P_{d e f}^{{dc}}=z^{dc}_{d} \cdot (g_d \cdot U_e^{\mathrm{dc}} \cdot (U_e^{dc}-U_f^{dc})),  \quad \forall d e f \in \dctopology \cup \dctopologyrev$ & \label{P_dc_def} \modeltwo \\
        $P_c^{cv,dc}= z^{cv}_{c} \cdot ( U_e^{dc} \cdot I_c^{cv,dc}),  \quad \forall c i e \in \mathcal{T}^{cv}$ & \label{P_cv} \modeltwo \\
        $P_c^{cv, loss}=z^{cv}_{c}\cdot a_c+b_c \cdot I_c^{cv,dc} + c_c \cdot (I_c^{cv,dc})^2,  \quad \forall c i e \in \convertertopology$ & \label{P_cv_loss} \modeltwo \\
        $-z^{cv}_{c} \cdot \underline{I}_c^{cv,dc} \leq I_c^{cv,dc} \leq z^{cv}_{c} \cdot \overline{I}_c^{cv,dc} \forall c i e \in \mathcal{T}^{\mathrm{cv}}$ & \label{I_cv_mag} \modeltwo \\
        $ -z^{dc}_{d} \cdot \underline{P}_{def} \leq P_{d e f} \leq z^{dc}_{d} \cdot \overline{P}_{def},  \quad \forall d e f \in \dctopology \cup \dctopologyrev$ \label{P_dc_dfe} & \modeltwo  \\
        $ -z^{cv}_{c}\cdot \underline{P}_c^{\mathrm{cv}, \mathrm{ac}} \leq P_c^{\mathrm{cv}, \mathrm{ac}} \leq z^{cv}_{c}\cdot \overline{P}_c^{\mathrm{cv}, \mathrm{ac}},  \quad \forall c i e \in \mathcal{T}^{\mathrm{cv}}$ \label{P_cv_ac} & \modeltwo  \\
        $ -z^{cv}_{c}\cdot \underline{P}_c^{c \mathrm{v}, \mathrm{dc}} \leq P_c^{\mathrm{cv}, \mathrm{dc}} \leq z^{cv}_{c}\cdot \overline{P}_c^{cv, dc},  \quad \forall c i e \in \mathcal{T}^{\mathrm{cv}}$ \label{P_cv_dc} & \modeltwo \\
        \hline
	\end{tabular}
    }
\end{table}

\subsubsection{AC and DC Busbar Splitting model} \label{sec:BS}
Fig. \ref{fig:switch_model} shows a simple model of the AC ${sw_{\upsilon mi}^{ac}}$ and DC ${sw_{\xi re}^{dc}}$ switches to perform BS on AC and DC busbars. If the switch is closed, the parameters of the two buses at their extremes are identical. Thus, they are considered a single bus. If the switch is open, the two parts are electrically distant even if they are physically close. As a result, their voltage magnitudes and angles can differ. Note that we use the term ``switch" to refer to every switching unit involved in the described busbar splitting model. The modeled switches can perform the operations of an actual switch, a disconnector, or a circuit breaker. %Further work will extend the model to differentiate the switching units mentioned above by including a set of distinctive constraints for each of them. As a result, grid protection methodologies for hybrid AC/DC grids~\cite{DC_protections} will be included in steady-state simulations.

\begin{figure}[h!]
		\centering
		\includegraphics[width = 0.5\linewidth]{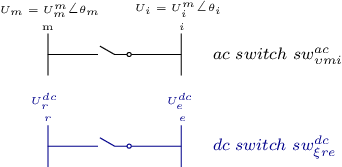}
		\caption{AC and DC switch models for busbar splitting.} 
		\label{fig:switch_model}
\end{figure}

%\begin{figure*}[h!]
%    \centering
%    \subfloat{{\includegraphics[width=6.0cm]%{Figures/busbar_base_model_crop.pdf} }}%
%    \qquad
%    \subfloat{{\includegraphics[width=11.0cm]%{Figures/busbar_split_model_crop.pdf}}}%
%    \caption{Busbar splitting representation with AC and DC %switches for AC and DC busbars. Each grid element %originally connected to the split busbars is attached to %an auxiliary bus and linked to each part of the split %busbar through a switch.}%
%    \label{fig:busbar_splitting_visualization}%
%\end{figure*}

Fig. \ref{fig:busbar_splitting_visualization} shows the BS methodology for AC and DC busbars. On the left-hand side, every grid element connected to the AC busbar $i$ is removed from the busbar and connected to a new AC bus $m$. Switches ${sw_{\upsilon mi}^{ac}}$ and ${sw_{\upsilon mi'}^{ac}}$ connect node $m$, to either part ($i$ or $i'$) of the split busbar $i$, and the busbar coupler ${sw_{\upsilon ii'}^{ac}}$ links the two parts of busbar $i$.
If the busbar coupler ${sw_{\upsilon ii'}^{ac}}$ is closed, the two parts of the AC busbar are connected and AC buses $i$ and $i'$ have the same complex voltage values. On the right-hand side of Fig. \ref{fig:busbar_splitting_visualization}, the DC busbar $e$ is split with the same methodology as the AC part.
The DC switches ${sw_{\xi re}^{dc}}$ and ${sw_{\xi re'}^{dc}}$ link the new bus $r$ to the two parts of the split DC busbar $e$ and $e'$, with the DC busbar coupler ${sw_{\xi ee'}^{dc}}$ connecting them.

\begin{figure*}[h!]
		\centering
		\includegraphics[width = 0.95\textwidth]{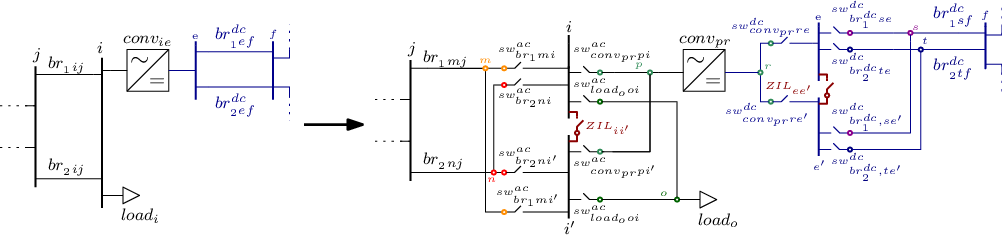}
		\caption{Busbar splitting representation with AC and DC switches for AC and DC busbars. Each grid element originally connected to the split busbars is attached to an auxiliary bus and linked to each part of the split busbar through a switch.} 
		\label{fig:busbar_splitting_visualization}
\end{figure*}

The total number of switches ($\acswitches + \dcswitches$) added to the model equals $\acswitches + \dcswitches = (\sum_{b=1}^{B} 2*n_{b}) + B$ where $B$ is the number of busbars being split and $n_{b}$ is the number of grid elements connected to each busbar. See sections \ref{AC_switch} and \ref{DC_switch} for the equations describing the AC and DC switches.

Note that the proposed BS methodology is flexible in terms of the grid elements which can be connected to either part of the split busbar. From a planning perspective, this flexibility helps to assess the best possible configuration for a given network. In a practical setting, certain grid elements are likely always connected to a given part of a (split) busbar. These ``fixed" elements result in fewer binary variables in the model.

\subsubsection{AC switch model} \label{AC_switch}
The AC switches ${sw_{\upsilon mi}^{ac}}$, ${sw_{\upsilon mi'}^{ac}}$ and ${sw_{ZILii'}^{ac}}$ are subject to (M2.14)-(M2.19) \cite{PowerModels2018}.
(M2.14) and (M2.15) deal with the voltage angles and the voltage magnitudes of the two buses at the extremes of each AC switch $sw^{ac}_{\upsilon mi}$. Moreover, (M2.16) and (M2.17) limit the active and reactive powers of the AC switch to its maximum $\overline{P}^{sw,ac}_{\upsilon mi}$,$\overline{Q}^{sw,ac}_{\upsilon mi}$ and minimum values $\underline{P}^{sw,ac}_{\upsilon mi}$,$\underline{Q}^{sw,ac}_{\upsilon mi}$, whereas (M2.18) defines the relation between both powers and the maximum apparent power of the switch $S^{sw,ac}_{\upsilon mi}$. Lastly, (M2.19) is an ``exclusivity" constraint including the switches connecting each grid element to the split busbar. It is either an equality ($=$ 1) or inequality ($\leq$ 1) constraint depending on whether OTS is performed or not. If OTS and BS are both allowed in the same optimization problem, (M2.19) is an inequality constraint and switches $sw^{ac}_{\upsilon mi}$ or $sw^{ac}_{\upsilon mi'}$ are both allowed to be open. As a result, the grid element is not reconnected to the split busbar $i$. If (M2.19) is an equality constraint, each grid element decoupled from the original busbar $i$ needs to be reconnected to one part of the split busbar, and one of the two switches $sw^{ac}_{\upsilon mi}$ or $sw^{ac}_{\upsilon mi'}$ must be closed. The possible switching states allowed by the ``exclusivity" constraint are represented in Figure~\ref{fig:busbar_splitting_excl_constraint}, where $1$ indicates that the switch is closed, $0$ that the switch is open. Note that constraint (M2.20) imposes that if the busbar coupler is closed, i.e. busbar splitting is not performed, the switch $sw^{ac}_{\upsilon mi}$ connecting the network element to the original busbar will always be closed.

\begin{figure}[h]
		\centering
		\includegraphics[width = 0.65\linewidth]{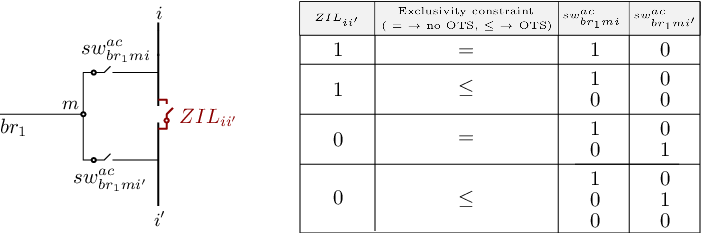}
		\caption{Possible configurations of the exclusivity constraint for each network element connected to a busbar being split.} 
		\label{fig:busbar_splitting_excl_constraint}
\end{figure}

\subsubsection{DC switch model} \label{DC_switch}
As DC grids do not have voltage angles and reactive power, the set of constraints (M2.21)-(M2.24) for DC switches include only the voltage magnitudes (M2.20) and active power (M2.21). Similarly to the previous ``exclusivity" constraint, (M2.19), (M2.22) are inequalities ($\leq$ 1) when OTS is allowed, and equality constraints otherwise. (M2.24) connects the network element to the original busbar if busbar splitting is not performed.

\begin{table}[h!]
	% increase table row spacing, adjust to taste
	\renewcommand{\arraystretch}{1.0}
	\centering
	%		\caption{Nonlinear Nonconvex  grid TNEP problem}
	\label{tb:Model212}
        {\fontsize{8pt}{11pt}\selectfont
	\begin{tabular}{m{40em} l}
		\hline
        Model 2.1.2: Busbar splitting & \\
        \hline
        \textbf{AC Busbar Splitting}&\\
        $z^{sw,ac}_{\upsilon mi} \cdot \theta_{m} = z^{sw,ac}_{\upsilon mi} \cdot \theta_{i},  \quad \forall \upsilon mi \in \acswitchtopology \cup \acZILtopology, \label{U_ac_a_sw}$ & \modeltwo \\
        $z^{sw,ac}_{\upsilon mi} \cdot U^{m}_{m} = z^{sw,ac}_{\upsilon  mi} \cdot U^{m}_{i},  \quad \forall \upsilon mi \in \acswitchtopology \cup \acZILtopology, \label{U_ac_m_sw} $ & \modeltwo \\
        $z^{sw,ac}_{\upsilon  mi} \cdot \underline{P}^{sw,ac}_{\upsilon  mi} \leq P^{sw,ac}_{\upsilon  mi} \leq z^{sw,ac}_{\upsilon  mi} \cdot \overline{P}^{sw,ac}_{\upsilon mi},  \quad \forall \upsilon mi \in \acswitchtopology \cup \acZILtopology \label{P_ac_sw} $ & \modeltwo \\
        $z^{sw,ac}_{\upsilon mi} \cdot \underline{Q}^{sw,ac}_{\upsilon mi} \leq Q^{sw,ac}_{\upsilon mi} \leq z^{sw,ac}_{\upsilon mi} \cdot \overline{Q}^{sw,ac}_{\upsilon mi},  \quad \forall \upsilon mi \in \acswitchtopology \cup \acZILtopology \label{Q_ac_sw}$ & \modeltwo \\
        $({P^{sw,ac}_{\upsilon mi}})^2 + ({Q^{sw,ac}_{\upsilon mi}})^2 \leq z^{sw,ac}_{\upsilon mi}\cdot({\overline{S}^{sw,ac}_{\upsilon mi}})^2,  \quad \forall \upsilon mi \in \acswitchtopology \cup \acZILtopology, \label{S_ac_sw}$ & \modeltwo \\
        $z^{sw,ac}_{\upsilon mi} + z^{sw,ac}_{\kappa mi'} \leq 1,  \quad \forall (\upsilon mi, \kappa mi') \in \acswitchtopology  \label{z_mn_sw_ots}$ & \modeltwo  \\
        $z^{sw,ac}_{\kappa mi'} \leq (1 - z^{sw,ac}_{ZILii'}),  \quad \forall (\kappa mi', ZILii') \in \acswitchtopology \cup \acZILtopology \label{z_mn_sw_integer_cut}$ & \modeltwo  \\
        \\
        \textbf{DC Busbar Splitting}&\\
        $z^{sw,dc}_{\xi re} \cdot U^{dc}_{r} = z^{sw,dc}_{\xi re} \cdot U^{dc}_{e},  \quad \forall \xi re \in \dcswitchtopology \cup \dcZILtopology$ \label{U_dc_sw} & \modeltwo \\
        $z^{sw,dc}_{\xi re} \cdot \underline{P}^{sw,dc}_{re} \leq P^{sw,dc}_{re} \leq z^{sw,dc}_{\xi re} \cdot \overline{P}^{sw,dc}_{\xi re},  \quad \forall \xi re \in \dcswitchtopology \cup \dcZILtopology$ \label{P_dc_sw} & \modeltwo \\
        $z^{sw,dc}_{\xi re} + z^{sw,dc}_{\iota re'} \leq 1,  \quad \forall \xi re,\iota re' \in \dcswitchtopology$ \label{z_ef_sw_1_ots} & \modeltwo \\
        $z^{sw,dc}_{\iota re'} \leq (1 - z^{sw,dc}_{ZILee'}),  \quad \forall (\iota re', ZILee') \in \dcswitchtopology \cup \dcZILtopology \label{z_mn_sw_integer_cut}$ & \modeltwo  \\
        \hline
	\end{tabular}
    }
\end{table}

\subsubsection{AC-Busbar splitting model with AC and DC switches}
The AC-OPF formulation introduced in section \ref{sec:OPF} is refined in Model 2.2 by including the constraints related to the AC (M2.14)-(M2.20) and DC switches (M2.21)-(M2.24) and by including the switches in the AC (M2.26) and DC nodal power balance (M2.27). In addition, a penalty term $c^{ac}_{sw}, c^{dc}_{sw}$ is added for each AC ($\acZILtopology$) and DC ($\dcZILtopology$) busbar coupler, respectively, in (M2.25). This term is large enough to ensure that BS is performed only if there is an actual economic benefit for such an operation. If BS is applied only to the AC or DC part of the network, the original nodal power balance equation (M1.5) or (M1.15) can be used for the part that is not split.
\begin{table}[h]
	% increase table row spacing, adjust to taste
	\renewcommand{\arraystretch}{1.0}
    \fontsize{8pt}{11pt}\selectfont
	\centering
	%		\caption{Nonlinear Nonconvex  grid TNEP problem}
	\label{tb:Model1}
	\begin{tabular}{m{40em} l}
		\hline
		Model 2.2: AC-BS for hybrid AC/DC grids &\\
		\hline
        \textbf{Minimize}: & \\
        $\sum_{k \in G} c_{2k}\cdot {P_{k}^{g}}^{2} + c_{1k}\cdot P_{k}^{g} + c_{0k} + \sum_{ZILii' \in \acZILtopology} c^{ac}_{sw}\cdot z^{sw,ac}_{ZILii',t} + \sum_{ZILee' \in \dcZILtopology} c^{dc}_{sw}\cdot z^{sw,dc}_{ZILee',t}$ & \modeltwo \\
        \\
        \textbf{Subject to}: & \\
        (M1.2)-(M1.4), (M1.6)-(M1.14), (M1.16)-(M1.17), (M2.14)-(M2.22), & \\
        \\
        \textbf{AC bus power balance}&\\
        $\sum_{\substack{k \in \generators_i}} S^g_k + \sum_{\substack{l \in \loads_{i}}} S^m_l - \sum_{\substack{c \in \acdcconverters_{i}}} S^{cv,ac}_c - y_{i} \cdot |U_i|^2 + \sum_{\substack{\upsilon mi \in (\acswitchtopology_{i} \cup \acZILtopology_{i})}} {S^{sw,ac}_{\upsilon mi}} = \sum_{\substack{lij\in \actopology}} S^{ac}_{lij},  \quad \forall i \in \acnodes$ & \modeltwo \\ 
        \\
        \textbf{DC bus power balance}&\\
        $\sum_{\substack{k \in \dcgenerators_e}} P^{g,dc}_k + \sum_{\substack{c \in \acdcconverters_e}} P^{cv, dc}_c - \sum_{\substack{l \in \dcloads}} {P^{m,dc}_l} - y_{e} \cdot |U_e|^2 + \sum_{\substack{\xi re \in (\dcswitchtopology_{e} \cup \dcZILtopology_{e})}} {P^{sw,dc}_{\xi re}} = \sum_{\substack{def} \in \dctopology} P^{dc}_{def},  \quad \forall e \in \dcnodes$ & \modeltwo \\ 
        \hline
	\end{tabular}
\end{table}

\subsection{Big M reformulation of the AC and DC switch models} \label{sec:Reformulation}
The formulations for the AC and DC switches include equations with bilinear terms, namely (M2.14), (M2.15) and (M2.20). Using the big M formulation~\cite{Linear_optimization} mentioned in Section~\ref{sec:literature_review}, these nonlinear constraints are reformulated in the linear constraints shown in (M2.28 - M2.30). Standard values for $M_{\theta}$ ($2\pi$), $M_{m}$ ($1.0$), and $M_{dc}$ ($1.0$) are used for the difference between the voltage angles and magnitudes connected by each switch. Note that while recent work has focused on optimizing the Ms for the OTS problem~\cite{PINEDA2024110620}, selecting optimal M values is beyond the scope of this paper and will be addressed in future work. 
\begin{table}[h!]
	% increase table row spacing, adjust to taste
	\renewcommand{\arraystretch}{1.0}
    \fontsize{8pt}{11pt}\selectfont
	\centering
	%		\caption{Nonlinear Nonconvex  grid TNEP problem}
	\label{tb:Model1}
	\begin{tabular}{m{40em} l}
		\hline
		Model 2.3: Big M reformulation for Busbar Splitting &\\
		\hline
        \textbf{AC Busbar Splitting with big M reformulation}&\\
        $ - (1 - z^{sw,ac}_{\upsilon mi}) \cdot M_{\theta} \leq \theta_{m} - \theta_{i} \leq (1 - z^{sw,ac}_{\upsilon mi}) \cdot M_{\theta},  \quad \forall \upsilon mi \in \acswitchtopology \cup \acZILtopology$ \label{diff_leq_M_delta} & \modeltwo\\
        %$ - (1 - z^{sw,ac}_{\upsilon mi}) \cdot M_{\theta} \leq \theta_{i} - \theta_{m} \leq (1 - z^{sw,ac}_{\upsilon mi}) \cdot M_{\theta} \quad \forall \upsilon mi \in \acswitchtopology \cup \acZILtopology$ \label{diff_leq_M_delta_inv} & \modeltwo\\
        $ - (1 - z^{sw,ac}_{\upsilon mi}) \cdot M_{m} \leq U^{m}_{m} - U^{m}_{i} \leq (1 - z^{sw,ac}_{\upsilon mi}) \cdot M_{m}
        ,  \quad \forall \upsilon mi \in \acswitchtopology \cup \acZILtopology$ \label{diff_leq_M_m} & \modeltwo\\
        %$ - (1 - z^{sw,ac}_{\upsilon mi}) \cdot M_{m} \leq U^{m}_{i} - U^{m}_{m} \leq (1 - z^{sw,ac}_{\upsilon mi}) \cdot M_{m},\quad \forall \upsilon mi \in \acswitchtopology \cup \acZILtopology$ \label{diff_leq_M_m_inv} & \modeltwo \\
        \\
        \textbf{DC Busbar Splitting with big M reformulation}&\\
        $ - (1 - z^{sw,dc}_{\xi re}) \cdot M_{dc} \leq U^{dc}_{r} - U^{dc}_{e} \leq (1 - z^{sw,dc}_{\xi re}) \cdot M_{dc}
        ,  \quad \forall \xi re \in \dcswitchtopology \cup \dcZILtopology $ \label{diff_leq_M_m} & \modeltwo \\
        %$ - (1 - z^{sw,dc}_{\xi re}) \cdot M_{dc} \leq U^{dc}_{r} - U^{dc}_{e} \leq (1 - z^{sw,dc}_{\xi re}) \cdot M_{dc},  \forall \xi re \in \dcswitchtopology \cup \dcZILtopology$ \label{diff_leq_M_m_inv} & \modeltwo \\
        \hline
	\end{tabular}
\end{table}

\begin{table}[h!]
	% increase table row spacing, adjust to taste
	\renewcommand{\arraystretch}{1.0}
	\fontsize{8pt}{11pt}\selectfont
    \centering
	%		\caption{Nonlinear Nonconvex  grid TNEP problem}
	\label{tb:Model1}
	\begin{tabular}{m{40em} l}
		\hline
		Model 2.4: AC-BS for hybrid AC/DC grids with big M reformulation &\\
		\hline
        \textbf{Minimize}: (M2.25) & \\
        \textbf{Subject to}: & \\
        (M1.2)-(M1.4), (M1.6)-(M1.14), (M1.16)-(M1.17), & \\ (M2.16)-(M2.19), (M2.21)-(M2.24), (M2.26)-(M2.30)& \\
        \hline
	\end{tabular}
\end{table}

%\subsection{Mixed Integer, Second-Order Cone Programming (MISOCP) and Mixed Integer Linear Programming (MILP) reformulations of the AC-BS problem for hybrid AC/DC grids}\label{Convex_relaxations}
\subsection{Reformulations of the AC-BS problem for AC/DC grids}\label{Convex_relaxations}
\subsubsection{Second Order Cone (SOC) Relaxation}
Several variables introduced previously are modified to build a SOC problem~\cite{SOC} by lifting the variables to a higher dimensional space:
\begin{align}
    & (U_{i}^{m})^{2} \rightarrow W_{i}, \; (U_{j}^{m})^{2} \rightarrow W_{j}, \; U_{i}^{m} \cdot U_{j}^{m} \rightarrow W_{ij}, \notag \\ 
    & (U_{e}^{dc})^{2} \rightarrow W^{dc}_{e}, \; (U_{f}^{dc})^{2} \rightarrow W^{dc}_{f}, \; U_{e}^{dc}  \cdot U_{f}^{dc} \rightarrow W_{ef}^{dc}, \notag\\
    & (I^{dc}_{def})^{2} \rightarrow i^{sq,dc}_{def}, \;
    (I^{cv,dc}_{c})^{2} \rightarrow i^{sq,cv}_{c}, \; (U_{c}^{f,mag})^{2} \rightarrow W^{f}_{c} \; \notag
\end{align}
The lifted variables are used to convexify the non-convex constraints from the original AC-OPF formulation \cite{QC}, such as equations (M1.7), (M1.8) for the AC grid and (M1.16) for the DC grid. Given that the SOC formulation is a well-established relaxation in literature \cite{SOC}, the full formulation for the AC-OPF for hybrid AC/DC grids is not added to this paper. The interested reader can find it in~\cite{ergun_optimal_2019}.
\subsubsection{Quadratic Convex (QC) Relaxation}
The QC relaxation uses the previously-mentioned lifted variables $W$ and builds convex envelopes around the nonlinear terms \cite{QC_math} by using the McCormick relaxation \cite{McCormick}. The reader is referred to the work of Coffrin et al. ~\cite{QC_math} for the full formulation and description of the QC relaxation.  For the DC grid part, the developed Bus Injection SOC (SOC BIM) and QC relaxations are equivalent, as mentioned in  \cite{ergun_optimal_2019}.
\subsubsection{Linear Programming Approximation (LPAC)} \label{LPAC_approximation}

Unlike the DC linearization~\cite{DC_OPF}, the LPAC approximation~\cite{LPAC} presents voltage magnitudes and reactive power in addition to voltage phase angles. Moreover, a piecewise linear approximation of the cosine term in the AC branches equations and Taylor series for approximating the nonlinear terms are used. This paper is based on the Cold-Start model from \cite{LPAC} where the target voltages $\tilde{V}^{t} = 1$ are the starting values and the power flows are based on the $\phi_{i}$ voltage magnitudes change $\phi_{i} - \phi_{j}$ for the AC grid and $\phi_{e} - \phi_{f}$ for the DC grid. As for the SOC and QC relaxations, the full LPAC-OPF formulation is not included in this paper, but can be found in~\cite{LPAC}.

Note that all the reformulations are still non-convex by definition due to the binary variables, but the continuous parts of the equations are convexified.

\subsubsection{AC and DC switches model adaptations}
The big M reformulation of the AC and DC switches models described previously in (M2.27)-(M2.29) are modified in (M3.1)-(M3.8) to include the lifted variables $W$ (SOC, QC relaxations) and voltage magnitude change  $\phi$ (LPAC approximation) previously discussed in the section.
\begin{table}[h!]
	% increase table row spacing, adjust to taste
	\renewcommand{\arraystretch}{1.0}
	\fontsize{8pt}{11pt}\selectfont
    \centering
	%		\caption{Nonlinear Nonconvex  grid TNEP problem}
	\label{tb:Model1}
	\begin{tabular}{m{40em} l}
		\hline
		Model 3: Big M reformulation for Busbar Splitting with SOC, QC relaxations and LPAC approximation &\\
		\hline
        \textbf{Convex relaxations} & \\
        $ - (1 - z^{sw,ac}_{\upsilon mi}) \cdot M_{m} \leq W_{m} - W_{i} \leq (1 - z^{sw,ac}_{\upsilon mi}) \cdot M_{m},  \quad \forall \upsilon mi \in \acswitchtopology \cup \acZILtopology$ & \modelthree \\
        %$ - (1 - z^{sw,ac}_{\upsilon mi}) \cdot M_{m} \leq W_{i} - W_{m} \leq (1 - z^{sw,ac}_{\upsilon mi}) \cdot M_{m},  \quad \forall \upsilon mi \in \acswitchtopology \cup \acZILtopology $ & \modelthree \\
        $ - (1 - z^{sw,dc}_{\xi re}) \cdot M_{dc} \leq W^{dc}_{r} -  W^{dc}_{e} \leq (1 - z^{sw,dc}_{\xi re}) \cdot M_{dc}
        ,  \quad \forall \xi re \in \dcswitchtopology \cup            \dcZILtopology$ & \modelthree \\
        %$ - (1 - z^{sw,dc}_{\xi re}) \cdot M_{dc} \leq W^{dc}_{r} -            W^{dc}_{e} \leq (1 - z^{sw,dc}_{\xi re}) \cdot M_{dc},  \quad \forall \xi re \in \dcswitchtopology \cup \dcZILtopology $ & \modelthree \\
        \\
        \textbf{Linear Approximation} & \\
        $ - (1 - z^{sw,ac}_{\upsilon mi}) \cdot M_{m} \leq \phi_{m} - \phi_{i} \leq (1 - z^{sw,ac}_{\upsilon mi}) \cdot M_{m},  \quad \forall \upsilon mi \in \acswitchtopology \cup \acZILtopology $ & \modelthree \\
        %$ - (1 - z^{sw,ac}_{\upsilon mi}) \cdot M_{m} \leq \phi_{i} - \phi_{m} \leq (1 - z^{sw,ac}_{\upsilon mi}) \cdot M_{m}   ,  \quad \forall \upsilon mi \in \acswitchtopology \cup \acZILtopology$ & \modelthree \\
        $ - (1 - z^{sw,dc}_{\xi re}) \cdot M_{dc} \leq \phi^{dc}_{r} - \phi^{dc}_{e} \leq (1 - z^{sw,dc}_{\xi re}) \cdot M_{dc},  \quad \forall \xi re \in \dcswitchtopology \cup \dcZILtopology $ & \modelthree \\
        %$ - (1 - z^{sw,dc}_{\xi re}) \cdot M_{dc} \leq \phi^{dc}_{e} - \phi^{dc}_{r} \leq (1 - z^{sw,dc}_{\xi re}) \cdot M_{dc},  \quad \forall \xi re \in \dcswitchtopology \cup \dcZILtopology $ & \modelthree \\
        \hline
	\end{tabular}
\end{table}
To have the relaxed and approximated versions of Model 2.4, (M2.27-M2.29) are substituted by the constraints (M3.1-M.3.2) and (M3.3-M3.4). Similarly, the OPF equations from~\cite{ergun_optimal_2019} are replaced by the formulations from~\cite{QC} (QC), ~\cite{SOC} (SOC) and ~\cite{LPAC} (LPAC).

\section{Test cases} \label{sec:test_cases}
The OTS and BS models are tested on five multi-terminal hybrid AC/DC grids of different sizes, namely 5-buses (shown in Fig. \ref{fig:case_5_original}), 39-buses, 588-buses and 3120-buses from \cite{ergun_optimal_2019} and a 67-buses from~\cite{case67acdc}. The number of grid elements for each of them is summarised in Table~\ref{table:test_cases}. Compared to~\cite{ergun_optimal_2019}, an adapted 39-buses case is evaluated in low-load operating conditions.
\begin{table}[h!]
\caption{Multi-terminal hybrid AC/DC test cases used in this paper.}
\fontsize{8pt}{10pt}\selectfont
\centering
\begin{tabular}{l|ccccc}
\hline
Test case             & \# AC & \# DC  & \#AC/DC & \# AC  & \# DC  \\
            & \ buses & buses & conv & branches & branches \\
\hline
5-buses~\cite{ergun_optimal_2019} & 5           & 3           & 3             & 7              & 3                   \\
39-buses~\cite{ergun_optimal_2019}  & 39          & 10          & 10            & 46             & 12                   \\
67-buses~\cite{case67acdc} & 67          & 9           & 9             & 102            & 11                   \\
588-buses~\cite{ergun_optimal_2019}  & 588         & 7           & 7             & 686            & 8                    \\
3120-buses~\cite{ergun_optimal_2019}       & 3120        & 5           & 5             & 3693           & 5                   \\
\hline
\end{tabular}
\label{table:test_cases}
\end{table}

The 5-buses test case from Fig. \ref{fig:case_5_original} is used in section \ref{sec:results} for a practical visualization of the results, while the other, larger cases confirm the proposed models' scalability. The results presented in the next Section \ref{sec:results} are computed using a MacBook Pro with chip M1 Max and 32 GB of memory.
\begin{figure}[h!]
		\centering
		\includegraphics[scale=1.5]{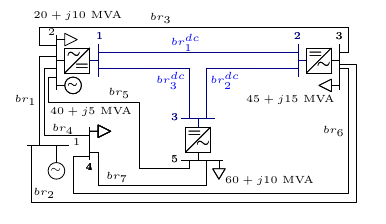}
		\caption{Case ``5-buses'' hybrid AC/DC. It consists of a 5-node AC grid with a 3-node meshed DC grid~\cite{ergun_optimal_2019}.} 
		\label{fig:case_5_original}
\end{figure}

\section{Results and discussion} \label{sec:results}
The AC-OPF, AC-OTS, and AC-BS simulations are performed using the Juniper optimizer~\cite{juniper}, combining Gurobi~\cite{Gurobi} and Ipopt~\cite{ipopt}. The QC and SOC relaxations and LPAC approximations are instead solved with Gurobi~\cite{Gurobi} given their MISOCP (SOC) and MIQCP (QC, LPAC) natures.

The results are organized in two parts. Firstly, the AC-OTS model with switchable grid elements in the AC, DC, and combined AC/DC parts is presented for the three smallest test cases mentioned in Section~\ref{sec:test_cases}. Due to the considerable amount of binary variables, the larger cases, 588-buses and 3120-buses, do not converge in an acceptable time for all the OTS models. Given the verbose and illustrative nature of this first part of the results, we limit the discussion to the exact MINLP formulation, for conciseness. The results for the other formulations are featured in the second set of results, described hereafter.

Secondly, the results of the AC-BS model with AC and DC switches are analyzed for the same test cases. The performed switching actions are shown on the case5-buses's grid topology introduced in Fig.~\ref{fig:case_5_original}.

\footnotetext[1]{All the models presented in the paper are publicly available at https://github.com/Electa-Git/PowerModelsTopologicalActions.jl.}

\subsection{Optimal Transmission Switching}
The results of the AC-OTS model with switchable grid elements in the AC, DC, and combined AC/DC parts are shown in Table \ref{table:OTS_results}.

\begin{table*}[h!]
\caption{Results for the OPF and AC-OTS models for the hybrid AC/DC test cases. \textbf{LO} stands for ``Lower Objective value”.}
\fontsize{9pt}{11pt}\selectfont
\centering
\begin{tabular}{l|l|ccccc}
\hline
Test case  & Model & Obj. value & \textbf{LO} wrt AC-OPF? & Time & \# Binary &  \# Branches and  \\
 & & [\$/h] &  Benefit wrt AC-OPF [\%] & [s] & variables & converters switched \\
\hline
5-buses         & AC-OPF             & 194.139  &  -  & 0.014            & -              & -                     \\
                & AC-OTS, AC         & 184.437  &  \ding{51}, 5.00  & 0.228            & 7              & 3 AC                  \\
                & AC-OTS, DC         & 194.139  &  \ding{55}, 0.00  & 0.353            & 6              & -                     \\
                & AC-OTS, AC \& DC   & 184.437  &  \ding{51}, 5.00  & 1.179           & 13             & 3 AC                  \\ 
                \hline
39-buses       & AC-OPF              & 2508.410  &  -  & 0.111            & -              & -                     \\
                & AC-OTS, AC         & 2508.410  &  \ding{55}, 0.00  & 9.801            & 46             & -                  \\
                & AC-OTS, DC         & 2501.240  &  \ding{51}, 0.29  & 11.582           & 22             & 5 DC, 6 conv dc  \\
                & AC-OTS, AC \& DC   & 2501.240  &  \ding{51}, 0.29  & 35.865          & 68             & 5 DC, 6 conv dc \\ 
                \hline
67-buses        & AC-OPF             & 122253.02 & -  & 0.136            & -              & -                     \\
                & {AC-OTS, AC}       & 122128.68 & \ding{51}, 0.10  & 1.712            & 102            & 12 AC                \\
                & AC-OTS, DC         & 122253.02 & \ding{55}, 0.00  & 4.774           & 20             & -                    \\
                & AC-OTS, AC \& DC   & 122128.68 & \ding{51}, 0.10  & 4.991           & 122            & 12 AC                \\
                \hline
\end{tabular}
\label{table:OTS_results}
\end{table*}

The AC-OTS reduces the generation costs compared to the AC-OPF for one operating point of all the feasible test cases, namely 5-buses,  39-buses, and 67-buses. The DC-OTS model with only DC switchable elements does not reduce the generation costs. Similarly to the DC-OTS, the AC/DC-OTS has the same total generation costs as the AC-OTS for cases 5-buses and 67-buses, as no DC branch nor AC/DC converter is de-energized in this simulation. Nevertheless, case 39-buses shows an additional reduction in the generation costs when 2 AC branches, 4 DC branches, and 8 AC/DC converters are switched off in the AC/DC-OTS. These results confirm how switching off DC branches and AC/DC converters can reduce the overall generation costs in AC/DC grids and show that the optimization model works as expected. However, the cost reduction from switching DC components is not significant in these test cases. From a planning perspective, using the model on real test cases can inform the system operator whether investing in DC switches and/or circuit breakers is worthwhile from a flexibility point of view.

A potential solution to the scalability issues of test cases 588-buses and 3120-buses would be the selection of a subset of switchable lines. In real grids, it is typically not possible to de-energize any line arbitrarily, thus this restriction would be natural. Alternatively, computational analyses could be performed a priori to select a convenient subset of the available switchable lines. 

\subsection{Busbar Splitting}
For ease of computation and lack of available data on the average number of busbars that can be split, we assume that only \textit{one} busbar in our test cases is ``splittable". Even though potentially all the substations in a test case can be split, this assumption allows us to prove the economic benefit brought by the proposed method, while keeping a low computational complexity of the model. Future work will elaborate on effective busbar selection methods to achieve bigger reductions in the generation costs due to the splitting of several selected busbars. 

 The busbar being split for each test case is selected by running the LPAC-BS model for each busbar separately and choosing the one leading to the highest reduction in generation cost compared to the AC-OPF. The LPAC-BS is used as it provides the best trade-off between AC-feasibility and computational time among the proposed methods. Given that the LPAC formulation is an approximation of the AC-OPF, there are no formal guarantees of the feasibility of its results in an AC formulation. Nevertheless, this Section shows that, for the selected test cases, the grid topology optimized by the LPAC-BS is AC-feasible and often leads to a reduction in total generation costs when tested with the AC-OPF formulation. Note that the AC-OPF model mentioned in this section is the nonlinear, non-convex model for hybrid AC/DC grids described previously in Model 1 of Section~\ref{sec:models}. 
 
 Only for the smallest test case 5-buses, the possibility of splitting more busbars is shown in Fig.~\ref{fig:comparison}, as it clarifies the working principle behind the proposed methodology.
 
\begin{figure*}[h!]
    \centering
    \includegraphics[width=\linewidth]{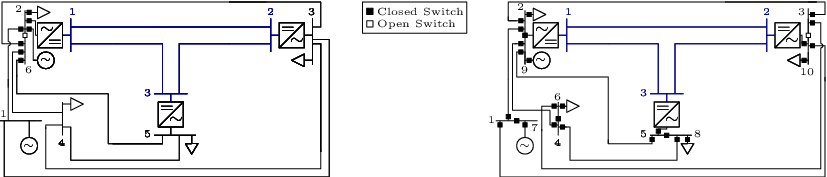}
    \caption{Optimized topologies for the AC-BS big M formulation applied to case 5-buses with AC busbar splitting possible respectively on AC busbar 2 (left) and all AC busbars (right).}
    \label{fig:comparison}
\end{figure*}
 
 On the left-hand side of Fig.~\ref{fig:comparison}, splitting busbar 2 leads to a reduction in the objective value for the AC-BS big M and LPAC-BS models. As displayed in Table~\ref{table:AC_BS_2}, the AC-BS big M (Model 2.4) leads to a lower objective value compared to the AC-OPF. Given the MINLP nature of the formulation, there is no guarantee that the solution is a global optimum, but the optimized topology still leads to a reduction in generation costs. While both convex relaxations keep the same configurations of the original test case, the LPAC approximation results in lower objective values than its OPF formulation. In addition, when an AC-feasibility check is run, the optimized topology is AC-feasible with a lower objective value compared to the original AC-OPF. Nevertheless, this AC-feasibility check's objective value is higher than the AC-BS big M MINLP formulation (Tables~\ref{table:AC_BS_2} and~\ref{table:AC_BS_2_4}) . This implies that the AC-BS big M formulation results in a higher reduction of generation costs compared to the LPAC-BS. In contrary, the LPAC-BS formulation is more than 100 times faster than the AC-BS big M model. Depending on the final application of the BS model, a trade-off should be made between a ``less'' optimal, yet considerably faster LPAC-BS formulation and an optimal but slower AC-BS big M formulation.

 These results are confirmed by Table~\ref{table:AC_BS_2_4} when all the AC busbars can be split in the same test case, with no maximum limit on the amount of busbars to be split. The final AC-BS big M topology is included on the right-hand side of Fig.~\ref{fig:comparison}. Even in this case, LPAC-BS proves to be significantly faster than the AC-BS big M while maintaining AC-feasibility. Its objective value is still higher than the AC-BS big M formulation, but lower than the AC-OPF. In addition, similarly to the previous case with one busbar being split, the AC-BS big M reduces further the total generation costs compared to the AC-OPF one, proving the value of splitting busbars. Note that while the LPAC-BS formulation keeps the topology optimized in Table~\ref{table:AC_BS_2}, the AC-BS big M finds a new optimal topology by splitting busbar 3. 
\begin{table*}[h!]
\fontsize{7.4pt}{10pt}\selectfont
\caption{Results for the AC-OPF and BS models for case 5-buses, AC busbar 2 split. \textbf{LO} stands for ``Lower Objective value”. \textit{fc} stands for feasibility check.}
\centering
\begin{tabular}{lcccccccc}
\hline
 Model          & OPF Obj.     & BS Obj.       & Time & Binary  & Time AC-\textit{fc} [s], & AC feasible? & Benefit wrt & Total \\
 & value [\$/h] & value [\$/h] & [s] & variables & Obj. value [\$/h] & \textbf{LO} wrt AC-OPF?  & AC-OPF [\%] & time [s] \\ 
\hline
AC-OPF      & 194.139 & -       & 0.014  & -   & -               & \ding{51}, -          & -     & 0.014 \\
AC-BS big M & 194.139 & 184.289 & 12.986 & 15  & 0.013, 184.289  & \ding{51}, \ding{51}  & 5.07 & 12.999 \\
SOC-BS      & 183.763 & 183.763 & 0.157  & 15  & 0.022, 194.139 & \ding{51}, \ding{55} & -    & 0.179  \\
QC-BS       & 183.761 & 183.761 & 0.287  & 15  & 0.023, 194.139 & \ding{51}, \ding{55} & -    & 0.310  \\
LPAC-BS     & 183.924 & 180.909 & 0.102  & 15  & 0.015, 186.349 & \ding{51}, \ding{51}   & 4.01 & 0.117  \\
\hline
\end{tabular}
\label{table:AC_BS_2}
\end{table*}

\begin{table*}[h!]
\fontsize{7.4pt}{10pt}\selectfont
\caption{Results for the AC-OPF and BS models for case 5-buses, all AC busbars can be split. \textbf{LO} stands for ``Lower Objective value”. \textit{fc} stands for feasibility check.}
\centering
\begin{tabular}{lcccccccc}
\hline
 Model          & OPF Obj.     & BS Obj.       & Time & Binary  & Time AC-\textit{fc} [s], & AC feasible? & Benefit wrt & Total \\
 & value [\$/h] & value [\$/h] & [s] & variables & Obj. value [\$/h] & \textbf{LO} wrt AC-OPF?  & AC-OPF [\%] & time [s] \\ 
\hline
AC-OPF  &    194.139     & -  & 0.014 & -  & - & \ding{51}, -                                 & - & 0.014 \\
AC-BS big M & 194.139 & 183.972   & 232.377 & 51  & 0.019, 183.972 & \ding{51}, \ding{51}   & 5.24  & 232.396  \\
SOC-BS   &   183.763  & 183.763   & 0.301  &  51  & 0.023, 194.139 & \ding{51}, \ding{55} & - & 0.324   \\
QC-BS   &    183.761   & 183.761  & 0.331  &  51  & 0.024, 194.139 & \ding{51}, \ding{55} & - & 0.355   \\
LPAC-BS  &   183.924 & 180.909  & 0.453    & 51   & 0.019, 186.349 & \ding{51}, \ding{51}   & 4.01 & 0.472   \\
\hline
\end{tabular}
\label{table:AC_BS_2_4}
\end{table*}
When extended to a larger network, the LPAC-BS is still the fastest model. In Table~\ref{table:AC_DC_BS_bigger}, the BS results are shown for several test cases with AC and DC busbars being split. All the AC-BS big M, SOC-BS, QC-BS, and LPAC-BS followed the same trends as Table~\ref{table:AC_BS_2} and Table~\ref{table:AC_BS_2_4}.  Since the convex relaxations do not lead to significant decreases in the total generation costs, they are not included in the remainder of the section. 

In Table~\ref{table:AC_BS_2_4}, splitting AC busbars leads to lower generation costs for all the AC-BS big M test cases. Besides case 39-buses, the LPAC-BS model is AC-feasible and has lower generation costs than the AC-OPF model in all test cases. Furthermore, its computational time is 10 to more than 200 times lower than the AC-BS big M model. These results suggest that there are clear benefits in terms of cost savings associated with BS and that the LPAC-BS model can provide an AC-feasible solution in a lower time compared to the AC-BS model.

Splitting DC busbars leads to better objective values for all the test cases for the AC-BS big M formulation. However, only case 39-buses is AC-feasible and has lower generation costs than the AC-OPF when the topology optimized with the LPAC-BS formulation is tested for AC-feasibility. In the other test cases, DC busbar splitting does not lead to a reduction in the total generation costs. Considering that not all test cases have an extended multi-terminal DC grid, these results confirm the potential benefit of splitting DC busbars and will be further investigated with more relevant test cases. In addition, future work will include AC/DC converter control and protection strategies in the developed models to better represent the operations of DC grids. Note that, even if OTS is allowed during BS, no grid elements are disconnected from the split busbars in any simulation.

\begin{table*}[h!]
\caption{Results for the AC-BS big M and LPAC-BS models for AC and DC busbar splitting in several hybrid AC/DC  
test cases. \textbf{LO} stands for ``Lower Objective value”.}
\fontsize{7pt}{10pt}\selectfont
\centering
\begin{tabular}{l|c|cccc|cccc}
\cline{3-10}
 \multicolumn{2}{c|}{} &  \multicolumn{4}{c|}{AC busbar split} &  \multicolumn{4}{c}{DC busbar split} \\
\hline
Test case,         & OPF Obj. & Split & BS Obj. & AC feasible?  & Time &  Split & BS Obj. & AC feasible? & Time \\
 Model &  value & busbar, & value & \textbf{LO} wrt & [s] & busbar, & value &  \textbf{LO} wrt & [s] \\
  & [\$/h]  & Binaries &  [\$/h] & AC-OPF? & & Binaries & &  AC-OPF? &  \\
\hline
39-buses, AC & 2508.410 & 2, 11 & 2507.830 & \ding{51}, \ding{51} & 30.3 & 4,  13 & 2508.390 & \ding{51}, \ding{51} & 32.1 \\
39-buses, LPAC    & 2469.120 & 2, 11 & 2468.200 & \ding{51}, \ding{55} & 0.5 & 4, 13 & 2468.870 & \ding{51}, \ding{51} & 1.3 \\
\hline
67-buses, AC & 122.253 & 43, 13 & 122.232 & \ding{51}, \ding{51} & 118.1 & 2, 7 & 118.125 & \ding{51}, \ding{51} & 43.2 \\
67-buses, LPAC   & 120.954 & 43, 13 & 120.942 & \ding{51}, \ding{51} & 0.5 & 2, 7 & 120.954 & \ding{51}, \ding{55} & 2.8  \\
\hline
588-buses, AC & 378.635 & 353, 19 & 376.523 & \ding{51}, \ding{51} & 958.1  & 4, 7 & 378.626 & \ding{51}, \ding{51} & 85.3 \\
588-buses, LPAC   & 372.198 & 353, 19 & 370.248 & \ding{51}, \ding{51} & 58.8   & 4, 7 & 372.198 & \ding{51}, \ding{55} & 18.2 \\
\hline
3120-buses, AC & 214.440 & 38, 21 & 214.231 & \ding{51}, \ding{51} & 12560.0 & 5, 7 & 214.263 & \ding{51}, \ding{51} & 266.7 \\
3120-buses, LPAC  & 211.906 & 38, 21 & 211.154 & \ding{51}, \ding{51} & 534.0   & 5, 7 & 211.906 & \ding{51}, \ding{55} & 95.6 \\
           \hline
\end{tabular}
\label{table:AC_DC_BS_bigger}
\end{table*}

% LPAC-BS    & 372.198 & 353 & 19 & 370.242 & \ding{51}, \ding{51} (value same $\uparrow$) & 52   & 4 & 7 & 372.198 & \ding{51}, \ding{55} & 18.0 \\

%Finally, Table~\ref{table:number_of_variables} shows the number of constraints, binary and continuous variables in the different models, together with their computational time. \textcolor{red}{ADD FURTHER ANALYSIS}.

%\begin{table}[h!]
%\caption{Number of constraints, continuous and binary %variables for each busbar splitting model in the different %multi-terminal hybrid AC/DC test cases.\textcolor{red}{TO BE %UPDATED}}
%\fontsize{6.9pt}{10pt}\selectfont
%\centering
%\begin{tabular}{l|ccc}
%\hline
%Test             &  \# of continuous & \# of binary & \# of \\
%case             & variables & variables & constraints \\
%\hline
%5-buses, one busbar~\cite{ergun_optimal_2019}  & 162         & %15  \\
%5-buses, all busbars~\cite{ergun_optimal_2019} & 274         & %51  \\
%39-buses~\cite{ergun_optimal_2019}, AC split   & 10          & %10   \\
%67-buses~\cite{case67acdc}, AC split           & 9           & %9   \\
%588-buses~\cite{ergun_optimal_2019}, AC split  & 7           & %7 \\
%3120-buses~\cite{ergun_optimal_2019}, AC split & 5           & %5    \\
%\hline
%\end{tabular}
%\label{table:number_of_variables}
%\end{table}

\section{Conclusion and future work}
This paper presents the first implementation of a combined optimal transmission switching and busbar splitting optimization model for hybrid AC/DC grids, in which the switching actions can be performed on both the AC and DC grid parts, jointly or separately, to reduce the total generation costs. Results from test cases of different sizes show that such actions lead to reduced generation costs compared to the original grid topology. %a ``plain" AC/DC-OPF on a given topology. 
While the benefits are test case-dependent, this reduction in generation costs confirms the effectiveness and potential of the proposed methodology. As future renewable-dominated AC/DC grids may increasingly be prone to congestion, resorting to topological actions guarantees a more effective network utilization at a lower cost compared to generation redispatch. 

The proposed BS models compare exact, non-convex power flow constraints and their convex relaxations and linear approximations. While all test cases in this paper solve in an acceptable time, a high computational time is expected in real-life test cases due to the increased number of variables. In particular, if multiple busbars are allowed to split, the guarantee of optimal or simply feasible results for the MINLP AC formulations needs to be further refined. Nevertheless, as shown in the paper, the MILP-LPAC formulation provides an excellent trade-off between speed, optimality, and feasibility, which provides AC-feasible results in a substantially lower computational time compared to the MINLP problem. 

Future work will include the development of methods (e.g., sensitivity analyses or metrics) to a priori shortlist the most ``promising" busbar candidates to be split. In addition, given the expected large RES penetration in the hybrid AC/DC grids of the future, the model will be extended to include RES forecasted uncertainty via stochastic optimization. Preventive N-1 security requirements will be added to the model to guarantee that the optimized topologies of the presented test cases are secure against contingencies taking place in the grid.

Lastly, the authors will contribute to the creation of real-life inspired test cases for AC/DC grids, which will allow for calculating the potential of topological actions in a more realistic fashion. This is expected, a.o., to result in more significant differences in the cost implied by different topologies, which does not fully emerge with the presently available synthetic test cases. % and better represent the multi-terminal DC grids in general.

\section*{Acknowledgement}
This paper has received support from the Belgian
Energy Transition Fund, FOD Economy, project DIRECTIONS.

\bibliographystyle{IEEEtran}
\bibliography{Bibliography}

\end{document}